\documentclass[letterpaper]{article} 
\usepackage{aaai25}  
\usepackage{times}  
\usepackage{helvet}  
\usepackage{courier}  
\usepackage[hyphens]{url}  
\usepackage{graphicx} 
\usepackage{natbib}  
\usepackage{caption} 
\usepackage{algorithm}
\usepackage{algorithmic}
\usepackage{amsthm,amsmath,amssymb,amsfonts}
\usepackage{color}
\usepackage{makecell}

\usepackage{mathrsfs}
\usepackage{array}

\usepackage{newfloat}
\usepackage{listings}
\usepackage{tabularx}
\usepackage{multirow}
\usepackage{booktabs}
\DeclareCaptionStyle{ruled}{labelfont=normalfont,labelsep=colon,strut=off} 
\floatstyle{ruled}
\newfloat{listing}{tb}{lst}{}
\floatname{listing}{Listing}
\title{AdaDPCC: Adaptive Rate Control and Rate-Distortion-Complexity Optimization for Dynamic Point Cloud Compression}
\author{
    Chenhao Zhang\textsuperscript{\rm 1}, 
    Wei Gao\textsuperscript{\rm 1,2}\thanks{Corresponding Author. }\\
}
\affiliations{
    \textsuperscript{\rm 1}Guangdong Provincial Key Laboratory of Ultra High Definition Immersive Media Technology, Shenzhen Graduate School, Peking University, Shenzhen 518055, China \\
    \textsuperscript{\rm 2}Peng Cheng Laboratory, Shenzhen 518055, China\\
    chenhaozhang@stu.pku.edu.cn, 
    gaowei262@pku.edu.cn
}

\begin{document}

\maketitle

\begin{abstract}
Dynamic point cloud compression (DPCC) is crucial in applications like autonomous driving and AR/VR. Current compression methods face challenges with complexity management and rate control. This paper introduces a novel dynamic coding framework that supports variable bitrate and computational complexities. Our approach includes a slimmable framework with multiple coding routes, allowing for efficient Rate-Distortion-Complexity Optimization (RDCO) within a single model. To address data sparsity in inter-frame prediction, we propose the coarse-to-fine motion estimation and compensation module that deconstructs geometric information while expanding the perceptive field. Additionally, we propose a precise rate control module that content-adaptively navigates point cloud frames through various coding routes to meet target bitrates. The experimental results demonstrate that our approach reduces the average BD-Rate by 5.81\% and improves the BD-PSNR by 0.42 dB compared to the state-of-the-art method, while keeping the average bitrate error at 0.40\%. Moreover, the average coding time is reduced by up to 44.6\% compared to D-DPCC, underscoring its efficiency in real-time and bitrate-constrained DPCC scenarios. Our code is available at
\url{https://git.openi.org.cn/OpenPointCloud/Ada_DPCC}.
\end{abstract}

%

\section{Introduction}
Dynamic point clouds, composed of sequences of point cloud frames, play crucial roles in applications such as robot sensing, autonomous driving, and AR/VR. These immersive technologies demand the real-time rendering of vast amounts of 3D data, posing significant challenges for data storage and transmission. To facilitate seamless and high-quality VR experiences, a robust yet flexible dynamic point cloud compression (DPCC) algorithm is indispensable, ensuring adaptable solutions under constraints of bandwidth and computational resources.

According to source coding theory \citep{2001_TC}, the essence of compression is the maximal elimination of source redundancy. In this context, existing DPCC efforts primarily focus on reducing spatial and temporal redundancies. To exploit spatial redundancy, current methods mainly utilize Variational Autoencoders to progressively downsample and aggregate spatial information into a latent space \citep{2019_AELGC,2021_PCGCv2,2022_DDPCC,2023_KABM_DPCC}. This results in a less-correlated latent representation, which is subsequently quantized and entropy-coded using established probability models \citep{2018_JointPrior,2018_VariationalHyper}. A significant limitation of these approaches is that they are optimized to a specific Rate-Distortion (RD) trade-off within a single model, limiting their flexibility in various bitrate scenarios. Conversely, exploiting temporal redundancy depends heavily on the precision of inter-frame coding. Current strategies typically involve motion estimation and KNN-based motion compensation within latent space \citep{2022_DDPCC,2023_KABM_DPCC,2024_patch_DPCC}. However, the high sparsity of downsampled points obscures the correspondence between points across frames, which in turn limits the perceptive field of KNN-based methods. The coding time is also significantly extended due to the computationally intensive nature of the KNN algorithm. Furthermore, existing methods often fail to provide precise rate control, lacking both accurate rate estimation and variable rate solutions, which are crucial for adjusting coding quality under diverse bandwidth constraints.

To address these challenges, this paper introduces a novel dynamic coding framework tailored for varying bitrates and computational complexities. We propose a dynamic Variational Autoencoder-based framework to efficiently exploit spatial redundancy, which includes multiple coding routes. Each route takes up a partial of the overall architectural computational complexity, achieving a distinct RD trade-off. Through joint training of these routes, optimal RD performance is attained at each complexity level, thereby facilitating RDCO within a single model. For temporal redundancy, we present a coarse-to-fine inter-prediction module that initially estimates motion embeddings and subsequently deconstructs geometric information from the latent space to provide anchors for precise motion compensation. Compared to KNN-based methods, our approach significantly reduces inter-prediction time while expanding the receptive field, leading to improved RD performance. Additionally, to achieve low-latency yet precise rate control, we introduce a lightweight rate control module that predicts the bitrate for each route and selects the optimal coding route using a sliding window and bit allocation algorithm.

The contributions of our work are as follows:
\begin{itemize}
\item[$\bullet$]  To achieve RDCO in the realm of DPCC, our method dynamically allocates computational complexity, maintaining superior RD performance. This flexibility allows for variable bitrate and adaptive computational resource allocation under bandwidth constraints.

\item[$\bullet$] To address data sparsity in inter-frame prediction, we introduce coarse-to-fine motion estimation and compensation, performing block-wise and voxel-wise motion estimation and anchor-based motion compensation.
\item[$\bullet$] To ensure precise rate control, the proposed method adaptively evaluates the bitrate of each route based on content and determines the optimal coding route through the sliding window and bit allocation algorithm.
\end{itemize}

\section{Related Work}
\subsection{Learning-based Point Cloud Compression}
The primary objective of learning-based PCC is to minimize the RD loss as:

\begin{equation}
\begin{aligned}
L=R+&\lambda D \\
= -\mathbb{E}(log_{2}P_{\tilde{y}}(\tilde{y}))& + \lambda d(x,\hat{x})
\label{eq:ratecontrolgoaldegenerate}
\end{aligned}
\end{equation}
where $\lambda$ is the Lagrange multiplier, $x$ is the input frame and $\hat{x}$ is the reconstructed frame, $P_{\tilde{y}}(\cdot)$ is the probability distribution of quantized latent representation $\tilde{y}$ and $d(\cdot,\cdot)$ is the criterion to measure the similarity between two frames. 

Learning-based PCC can be classified into point-based, voxel-based, and octree-based methods, each distinguished by the data format employed during the coding process.

\textbf{Point-based methods} utilize PointNet++-based auto-encoders to process point cloud coordinates and feature extraction. \citet{2019_3DPCGC} implemented Farthest Point Sampling (FPS) for downsampling coordinates and employed entropy coding for latent features. To mitigate the computational demands of FPS, \citet{2021_NGS} introduced Neural Graph Sampling with attention-based mechanisms on local graphs. \citet{2022_DPDPCC} used sub-point convolution to preserve density information during upsampling, and \citet{2024_COTPCC} innovated with a learnable sampler and Wasserstein distance for improved distribution fidelity. Despite their efficiency, point-based methods face challenges in memory management for large point clouds.

\textbf{Voxel-based methods} transform raw coordinates into gridded voxels, utilizing sparse convolution \citep{2019_ME,2023_TorchSparse} to optimize memory usage. \citet{2019_pcc_geo_cnn_v1} introduced 3D voxelized convolution within an auto-encoder framework, enhanced by a hyperprior framework in subsequent versions \citep{2020_pcc_geo_cnn_v2} to better explore spatial redundancy. \citet{2021_PCGCv1} incorporated Voxception-ResNet into their auto-encoder, with \citet{2021_PCGCv2} adding multi-scale upsampling to the decoder for improved transformation capabilities. \citet{2023_SparsePCGC} developed a unified framework for lossy and lossless compression, segmenting the coding process into scalable levels. Despite its effectiveness, the use of sparse convolution is restricted in handling context features, especially in sparse distributions from LiDAR scans.

\textbf{Octree-based methods} leverage the efficient octree structure to extract contextual information effectively. \citet{2020_OctSqueeze} initiated entropy coding of occupancy codes using ancestor nodes. \citet{2021_VoxelContextNet} refined the modeling of occupancy symbol probability distributions by leveraging local voxel contexts. \citet{2022_OctAttention} addressed gaps in resolution by integrating contexts from both ancestor and sibling nodes to minimize redundancy. To overcome the bottlenecks in serial decoding, \citet{2023_EHEM} introduced a hierarchical attention model, enhancing throughput while maintaining extensive context capture.

\subsection{Dynamic Point Cloud Compression}



DPCC primarily targets RD optimization across a Group of Frames (GoF), leveraging inter-frame correlation to enhance temporal context. Rule-based methods have been pivotal, with V-PCC \citep{2019_MPEG} encoding 3D frames onto 2D planes using traditional video coding techniques. Recent learning-based advancements have excelled in nonlinear transform coding \citep{2021_NTC} and probability estimation accuracy. Specifically, \citet{2024_IF_DPCC} integrated temporal information through sparse convolution and residual coding, resulting in superior RD metrics compared to V-PCC. However, this approach is limited by lacking explicit motion prediction. \citet{2022_DDPCC} enhanced motion estimation with multi-scale modules, and \citet{2023_KABM_DPCC} innovated with a KNN-attention block-matching (KABM) module for precise feature aggregation. Furthermore, \citet{2024_patch_DPCC} divided point cloud frames into patches to better explore inter-frame redundancy, setting new RD performance benchmarks. Despite these advancements, the computational demand of the KNN method limits the $K$ value in inter-prediction, thereby impacting local context capture. Additionally, the effectiveness of residual coding relies heavily on the accuracy of inter-prediction, underscoring the need for richer context exploitation.

\subsection{Dynamic Neural Network}
Dynamic Neural Network (DNN) is a novel category that adapts its architecture \citep{2020_AdaConv,2024_DyFormer,zhang2024learned} and parameters \citep{2023_InternImage,2023_SnakeConv} based on varying input data, making them highly suitable for learning-based compression. Addressing the demands for adaptive coding, \citet{2021_SlimCAE} introduced a slimmable auto-encoder to execute distinct RD trade-offs at various scales. Additionally, \citet{2023_AdaNIC} segmented input images into patches, dynamically assigning model capacities to match the coding complexities of each patch. \citet{2023_slimmabledecoder} further leveraged a slimmable decoder to adjust decoding complexity. Despite these innovations, developing an efficient DNN-based coding network for precise rate control remains a major challenge, highlighting the need for advanced DNN solutions.

\section{Methodology}
\begin{figure*}[t]
\centering
\includegraphics[width=0.9\textwidth]{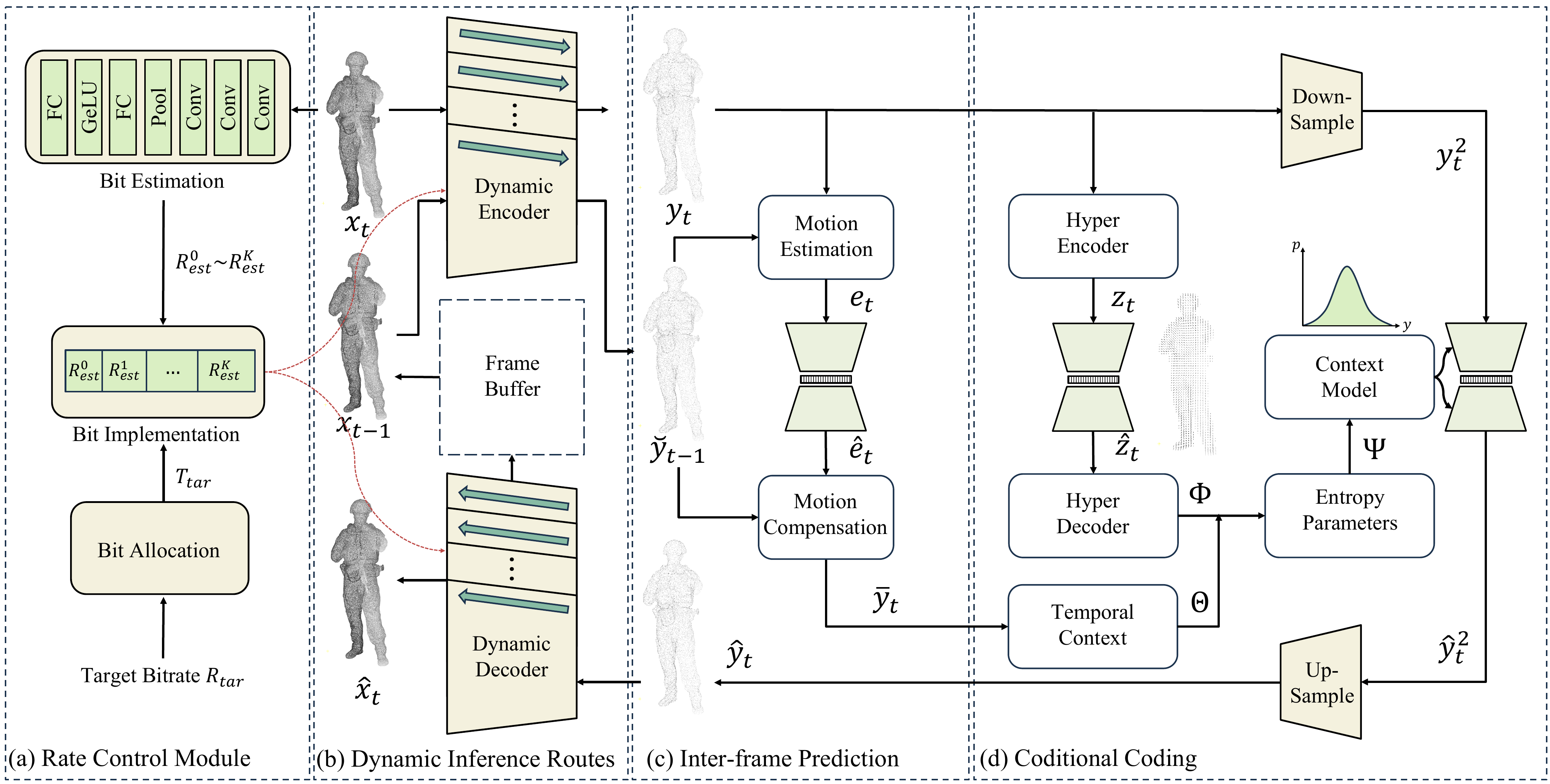} 
\caption{Overview of the proposed method: (a) Rate Control Module. It content-adaptively allocates coding routes for point cloud frames to meet target bitrates (b) Dynamic Inference Routes. The Dynamic Encoder/Decoder contains $K$ routes, each taking up partial of the total complexity and navigating to a unique RD trade-off. (c) Inter-frame Prediction. It exploits inter-frame redundancy with Motion Estimation and Motion Compensation modules. (d) Conditional Coding. It merges hyper-prior and temporal contexts to model the distribution of downsampled latent representation.}
\label{fig:framework}
\end{figure*}

\subsection{Overview}

To achieve RDCO and precise rate control, we propose a dynamic coding network with multiple coding routes. Specifically, to approach the target bitrate $R_{tar}$, the rate control module first estimates the coded bitrates and determines the optimal route $k$ for the current frame $x_t$. This frame is then encoded into latent representation $y_t$ via the $k$-th route of the dynamic encoder. Simultaneously, the reference frame $\hat{x}_{t-1}$ is encoded through the same route to generate $\breve{y}_{t-1}$. Coarse-to-fine motion estimation and compensation are applied between $y_t$ and $\breve{y}_{t-1}$, followed by conditional coding with the context of the compensated latent variable $\bar{y}_t$ and the hyper-prior $\hat{z}_t$. Finally, The reconstructed frame $\hat{x}_t$ is decoded via the same $k$-th route and buffered for subsequent inter-frame prediction. The overall architecture is shown in Figure \ref{fig:framework}.

\subsection{Dynamic Inference Routes}

\begin{figure}[t!]
\centering
\includegraphics[width=0.45\textwidth]{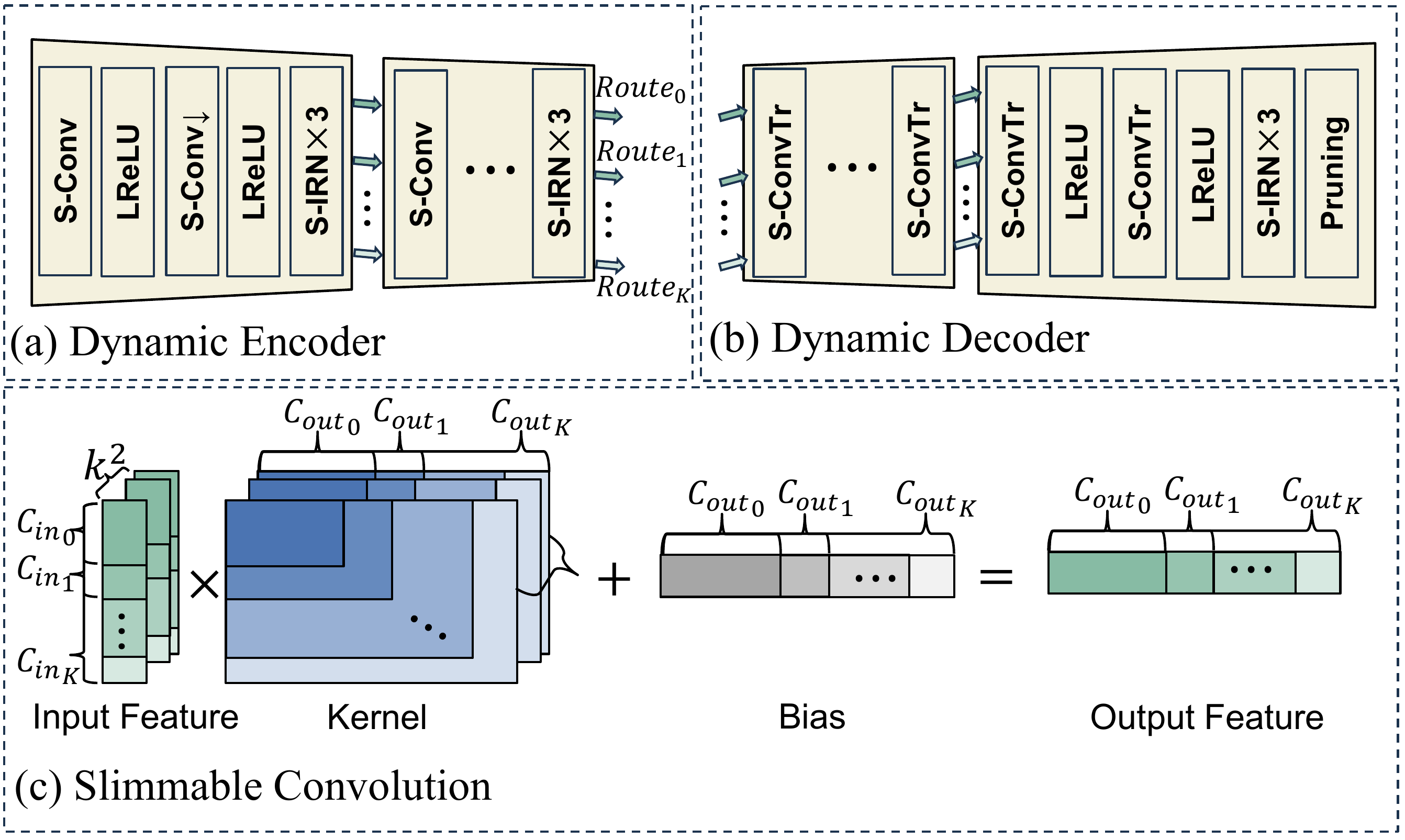}
\caption{Dynamic inference routes: (a) Dynamic Encoder with $K$ routes, each consists of two downsample blocks. (b) Dynamic Decoder with matching upsample blocks. ``S-X'' stands for `Slimmable X''. (c) Slimmable Convolution transferring variable input channels to output channels.}

\label{fig:dnn}
\end{figure}

It is noted that during the encoding of diverse point cloud contents, the neural network's active regions correlate with the contents' complexity, resulting in different active regions leading to varying bitrates. Therefore, our framework incorporates multiple coding routes, each contributing differently to computational complexity and resulting in distinct RD trade-offs. Specifically, the Dynamic Encoder $g_a(\cdot;\theta)$ and Dynamic Decoder $g_s(\cdot;\phi)$ utilize slimmable operators to dynamically adjust the channel dimensions of the input features, allowing the network to efficiently manage varying computational demands. These operators enable the division of the overall coding network (supernet) into several overlapping sub-networks (subnets), each handling a portion of the total computational load, as shown in Figure \ref{fig:dnn}. 

In the encoding process, the current point cloud frame $x_t = \{C_t, F_t\}$, where $C_t \in \mathbb{N}^{N \times 3}$ represents spatial coordinates and $F_t \in \mathbb{R}^{N \times D}$ denotes associated features with $D$ dimensions, is transformed into a $D_k$-dimensional latent representation $y_t = \{C_t^{'}, F_t^{'}\}$ via the selected route $k$:

\begin{equation}
y_t = g_a(x_t;\theta^{\leq k}) = \{C_t^{'}, F_t^{'}\},
\label{eq:encoder}
\end{equation}
where $\theta^{\leq k}$ denotes the parameters for the $k$-th route of the Encoder $g_a(\cdot;\theta)$. During encoding, the original point clouds are downsampled to sparser coordinates $C_t^{'}$ to encapsulate local geometric details in the feature space, enhancing the information capacity of the latent representation. Subsequently, $y_t$ is entropy coded under temporal and spatial contexts to maximize redundancy reduction, which will be elaborated in subsequent sections.

In the decoding process, the entropy-decoded latent representation $\hat{y}_t$ is transformed back to the reconstructed frame $\hat{x}_t = \{\hat{C}_t, F_t\}$ through the same route $k$:

\begin{equation}
\hat{x}_t = g_s(\hat{y}_t;\phi^{\leq k}) = \{\hat{C}_t, F_t\},
\label{eq:decoder}
\end{equation}
where $\phi^{\leq k}$ denotes the parameters for the $k$-th route of the Decoder $g_s(\cdot;\phi)$. Notably, in voxel space, features $F_t$ primarily serve as occupancy indicators and are uniformly set to all-one vectors.

As previously noted, routes with more parameters preserve more local geometric information, enhancing reconstruction quality at the expense of higher bitrate and computational complexity. However, it is inefficient to train each route independently due to the parameter interdependence among routes. To tackle this issue, we employ a joint training strategy, described in Algorithm \ref{alg:joint_training}. Initially, a supernet with comprehensive parameters is pre-trained using a high $\lambda$ to support high bitrate demands. Each subnet is then trained sequentially with decreasing $\lambda$ values, using the cumulative RD loss as the joint optimization criterion. Subsequently, the post-training process gradually reduces $\lambda_0$ for route $0$ to achieve the lowest acceptable bitrate.

\begin{algorithm}[!ht]
\caption{Joint Routes Training Strategy}
\label{alg:joint_training}
\begin{algorithmic}[1]
    \STATE\textbf{Input}: training iteration $N$, model $Model$, number of routes $K$, $\lambda$ list $[\lambda_{0},\lambda_{1},...,\lambda_{K-1}]$, train dataset $\chi_{train}$; 
    \STATE {pre-train the supernet with $\lambda_{K-1}$ for maximum bitrate.}
        \FOR {$j=1,2,\cdots,N$}
            \STATE $P_{input},P_{ref} \leftarrow \chi_{train}$
            \FOR {$k=K-2,K-3\cdots,0$}
                \STATE $R, D \leftarrow Model(P_{input}, P_{ref})$;
                \STATE $Loss \leftarrow Loss + R + \lambda_kD$;
            \ENDFOR
            \STATE update $Model$ parameters;
        \ENDFOR
    \STATE post-train by gradually decreasing $\lambda_0$
\end{algorithmic}
\end{algorithm}

Through joint routes training, the coding routes are collaboratively optimized to achieve optimal RD performance at each level of complexity. This approach allows the model to effectively manage computational complexity corresponding to variable bitrate, thereby supporting efficient RDCO within a single model.

\subsection{Coarse-to-fine Inter-frame Prediction}
\begin{figure*}[t!]
\centering
\includegraphics[width=1.0\textwidth]{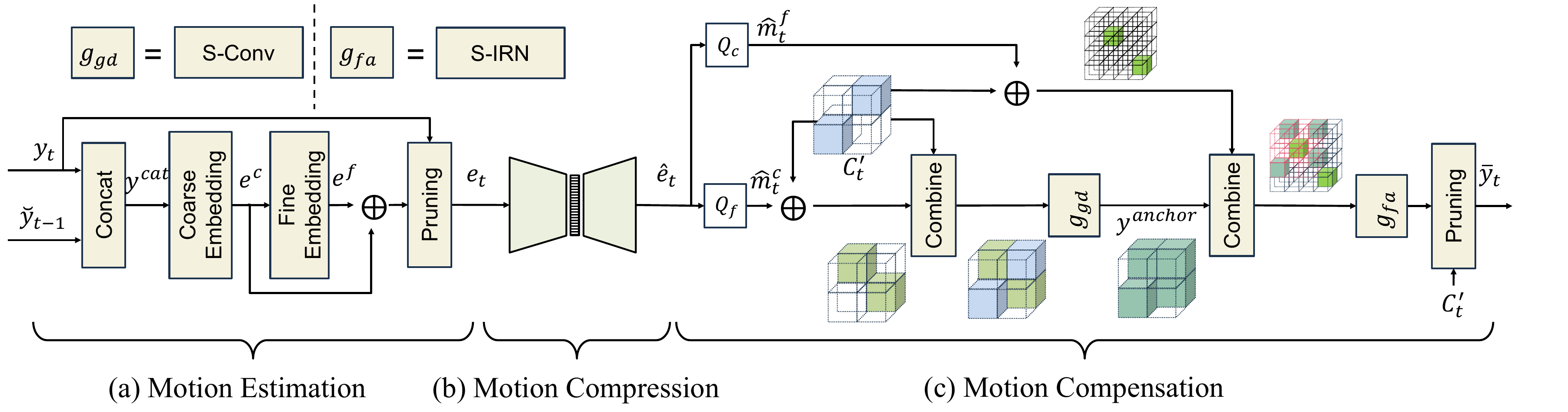}
\caption{Illustration of the coarse-to-fine inter-frame prediction.}
\label{fig:inter_prediction}
\end{figure*}

The accuracy of inter-frame prediction significantly impacts the quality of temporal context and, consequently, the RD performance. To address the challenges posed by large intervals in down-sampled coordinates and a limited perception field in inter-frame prediction, we introduce the Coarse-to-fine Inter-frame Prediction method, as shown in Figure \ref{fig:inter_prediction}.

\textbf{Motion estimation.} To achieve end-to-end optimization, inter-frame prediction is conducted in feature space. Temporal information from consecutive frames is fused by concatenating $y_t=\{C_t^{'},F_t^{'}\}$ and $\breve{y}_{t-1}=\{C_{t-1}^{'},F_{t-1}^{'}\}$ as:

\begin{equation}
C^{cat}=C_{t}^{'} \cup C_{t-1}^{'},
\label{eq:concat_coord}
\end{equation}

\begin{equation}
F^{cat}=
\begin{cases}
F_{t,m}^{'} \oplus F_{t-1,m}^{'} &  m\in C_{t}^{'} \cap C_{t-1}^{'} \\
F_{t,m}^{'} \oplus 0 &  m\in C_{t}^{'}, m\notin C_{t-1}^{'} \\
0 \oplus F_{t-1,m}^{'} &  m\notin C_{t}^{'}, m\in C_{t-1}^{'}
\end{cases},
\label{eq:concat_feat}
\end{equation}

\begin{equation}
y^{cat}=\{C^{cat},F^{cat}\},
\label{eq:concat}
\end{equation}
where $y^{cat}$ is the fused frame, $C^{cat}$ and $F^{cat}$ are the corresponding coordinates and features, $m$ represents any single point in $C^{cat}$, and $\oplus$ denotes channel-wise concatenation. The fused frame $y^{cat}$ is processed by the Motion Estimation module, which initially applies coarse-grained convolutions with the tensor stride of $2$. This is followed by fine-grained convolutions with the tensor stride of $1$, producing a fine-grained motion embedding $e_f$. To mitigate information loss from network depth, a residual branch directly integrates residual details with $e_f$. To ensure precise alignment with $y_t$, the coordinates of $e_t$ are pruned to match $C_t^{'}$, resulting in the final motion embedding $e_t=\{C_t^{'},m_t\}$. The coordinates of $e_t$ are encoded losslessly, while the 3D motion vector $m_t$ is quantized and entropy encoded into bitstream.

\textbf{Motion compensation.} The large intervals in down-sampled coordinates constrain the perceptive field, limiting the effectiveness of motion compensation. To address this, we introduce a two-shot motion compensation approach that first deconstructs geometric information from the feature space and then fuses neighboring features based on reconstructed anchors. This process begins by quantizing the decoded motion vector $\hat{m}_t$ to a $2\times$ down-sampled scale:

\begin{equation}
\hat{m}_t^{c} = \lfloor \hat{m}_t/2 \rfloor \times 2,
\label{eq:motion_course}
\end{equation}
where $\hat{m}_t^{c}$ represents the coarse-grained motion vector that captures block-wise motion vector. Subsequently, we warp the coordinates $C_t^{'}$ and combine them with $C_{t-1}^{'}$. To extract and deconstruct the geometric information from $\breve{y}_{t-1}$ into the warped voxels, the combined voxels are processed through the geometric deconstruction module $g_{gd}(\cdot)$:

\begin{equation}
y_t^{anchor} = g_{gd}(\{(\hat{m}_t^{c}+C_t^{'}),0\} \cup \{C_{t-1}^{'}, F_{t-1}^{'}\}),
\label{eq:anchor}
\end{equation}
where $y_t^{anchor}=\{C_t^{anchor}, F_t^{anchor}\}$ serves as the anchor, integrating geometric details from both voxel and feature spaces, thereby providing a rich and precise contextual anchor. For finer motion adjustments, we generate a fine-grained motion vector:

\begin{equation}
\hat{m}_t^{f} = \lfloor \hat{m}_t/1 \rfloor \times 1,
\label{eq:motion_fine}
\end{equation}
where $\hat{m}_t^{f}$ denotes fine-grained motion depicting voxel-wise motion vector. We then warp the coordinates of $C_t^{'}$ and merge them with $C_t^{anchor}$. These merged voxels are subsequently refined by the feature aggregation module $g_{fa}(\cdot)$ to align features from $y_t^{anchor}$ into the compensated latent representation $\bar{y}_t$:

\begin{equation}
\bar{y}_t = g_{fa}(\{(\hat{m}_t^{f}+C_t^{'}),0\} \cup y_t^{anchor}),
\label{eq:temporal_context}
\end{equation}
where $\bar{y}_t$ represents the predicted current frame in latent space, serving as the temporal context for conditional inter-frame coding. Through coarse-to-fine motion compensation, geometric information is reconstructed from the feature space, offering anchors that greatly enhance the accuracy of motion compensation.

\subsection{Conditional Inter-frame Coding}
\label{sec:conditional_coding}

To fully exploit geometric redundancy, we implement conditional inter-frame coding that first extracts context information from both temporal and spatial aspects. The overall context is generated as:







\begin{equation}
\Psi = g_{ep}(\Theta,\Phi;\phi_{ep}),
\label{eq:motion_fine}
\end{equation}
where $\Theta$ represents the temporal context, $\Phi$ denotes the hyper-prior context, and $\Psi$ serves as the overall context in conditional coding, with $g_{ep}(\cdot;\phi_{ep})$ being the entropy parameters network. The conditional probability $p_{\tilde{y}|\Psi}$ is modeled as a factorized Gaussian distribution integrated with soft quantization noise:

\begin{equation}
p_{\tilde{y}|\Psi}(\tilde{y}|\Psi)=\prod \limits_{i=0}(N(\mu_i,\sigma_i^2)\ast U(-\frac{1}{2},-\frac{1}{2}))(\tilde{y}_{t,i}),
\label{eq:motion_fine}
\end{equation}
where $U(-\frac{1}{2},\frac{1}{2})$ indicates the uniform distribution of soft quantization noise. Finally, the RD loss for the $k$-th route is:

\begin{equation}
\begin{aligned}
\mathscr{L}_k = R_{m}+R_{z}&+R_{y} + \lambda_{k}D \\
= -\mathbb{E}(log_{2}P_{\tilde{m}}(\tilde{m}))&-\mathbb{E}(log_{2}P_{\tilde{z}}(\tilde{z}))\\-\mathbb{E}(\sum_{i=1}log_2(p_{\tilde{y}_i}(\tilde{y}_i&|\Psi))) +\lambda_{k}d(x,\hat{x}),
\end{aligned}
\label{eq:loss}
\end{equation}
where $R_m$, $R_z$, and $R_y$ denote the rate losses for motion embedding, hyper-prior, and latent representation, respectively. Through end-to-end training, $R_m$ and $R_z$ contribute only a marginal portion to the final bitstream yet significantly enhance RD performance.

\begin{figure*}[t!]
\centering
\includegraphics[width=1.0\textwidth]{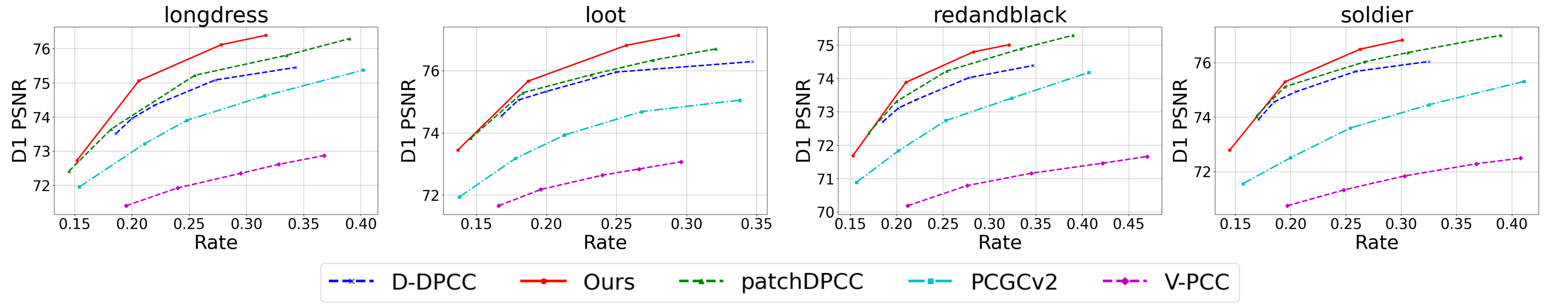}
\caption{D1-PSNR RD curves of different methods on 8iVFB dataset.}
\label{fig:RD_Curve}
\end{figure*}

\begin{table*}[t!]

  \centering
  \newcolumntype{C}{>{\centering\arraybackslash}X}
  \begin{tabularx}{\textwidth}{@{} p{0.105\textwidth} *{6}{C} @{}} 
    \toprule
    \multirow{2}{*}{Datasets}  & \multicolumn{4}{c}{D1 BD-Rate(\%)/BD-PSNR(dB)} \\
    \cline{2-5}
    &  D-DPCC & patchDPCC & PCGCv2 & V-PCC\\
    \midrule
    Longdress & -14.324/1.016 & -9.435/0.623 & -26.254/1.758 & -55.170/3.750\\
    Loot & -11.071/0.700 & -3.720/0.427 & -31.723/1.932 & -59.373/3.790\\
    Redandblack & -10.021/0.729 & -4.145/0.307 & -24.806/1.704 & -68.133/3.910\\
    Soldier & -9.485/0.683 & -5.941/0.338 & -59.764/2.747 & -67.082/4.935\\
    \textbf{Average}  & \textbf{-11.225/0.782} & \textbf{-5.810/0.423} &\textbf{-35.636/2.035} &\textbf{-62.439/4.096}\\
    \bottomrule
  \end{tabularx}
  \caption{RD performance comparison of different methods on 8iVFB dataset.}
  \label{tab:RDperformance}
\end{table*}

\subsection{Rate Control Module}
To achieve precise frame-level rate control, each frame's coding route is carefully selected to match the target bitrate $R_{tar}$. It is well-recognized that a frame's bitrate is closely linked to its spatial distribution and geometric representation. In the proposed framework, we integrate three key components to ensure accurate rate control: the Rate Estimation module, the Bit Allocation module, and the Bit Implementation module.

First, the Rate Estimation module estimates the bitrates $R_{est}^0\sim R_{est}^K$ for each coding route. Then, the Bit Allocation module employs a sliding window algorithm for frame-level bitrate allocation, determining the target bitrate $T_{tar}$ for the current frame:

\begin{align}
T_{tar} = \frac{R_{tar}\times(N_c+SW)-R_c}{SW},
\label{eq:SlidingWindow}
\end{align}
where $T_{tar}$ represents the target bitrate for the current frame, $N_c$ is the number of frames that have already been encoded, $R_c$ refers to the bits used by these frames, and $SW$ is the length of the sliding window.  This algorithm aims for bitrate stability across $SW$ frames, promoting consistent video quality. Notably, at the start of a GoF, the ``I'' frame is typically allocated a higher bitrate to enhance RD performance in subsequent frames, deviating from the regular allocation strategy to mitigate cumulative errors.

Finally, the Bit Implementation module identifies current route $i^*$ based on the current bit consumption. If the allocated bits exceed the consumed bits, the module selects the route with an estimated bitrate slightly above $T_{tar}$, or the highest available bitrate route if no such match is found. On the other hand, if the allocated bits are less than the consumed bits, the module chooses a route with an estimated bitrate slightly below $T_{tar}$, or the lowest available bitrate route if no suitable option is available. The bit implementation strategy is described as follows:
\begin{equation}
i^* = 
\begin{cases} 
\mathop{\arg\min}\limits_{i} \left( R_{est}^i - T_{tar} \right), & \text{if } R_{tar} \times N_c > R_c \\
\quad \text{s.t. } R_{est}^i > T_{tar} \\
\mathop{\arg\min}\limits_{i} \left( T_{tar} - R_{est}^i \right), & \text{if } R_{tar} \times N_c < R_c \\
\quad \text{s.t. } R_{est}^i < T_{tar}
\end{cases}
\label{eq:BitImplementation}
\end{equation}

\section{Experiments}

\begin{table}[t!]
  \centering
  \begin{tabular}{p{2.5cm}p{4cm}}
    \toprule
    \centering Methods & \centering\arraybackslash Coding Time (s)\\
    \midrule
    \centering D-DPCC & \centering\arraybackslash 1.28/1.16 \\
    \centering PCGCv2 & \centering\arraybackslash 1.09/0.84 \\
    \centering Ours(Route 3) & \centering\arraybackslash 1.07/1.16\\
    \centering Ours(Route 2) & \centering\arraybackslash 0.85/0.81\\
    \centering Ours(Route 1) & \centering\arraybackslash 0.77/0.75\\
    \centering Ours(Route 0) & \centering\arraybackslash 0.68/0.67\\
  \bottomrule
  \end{tabular}
  \caption{Average coding time comparison (encoding time/decoding time) on 8iVFB dataset.}
  \label{tab:complexity}
\end{table}

\subsection{Implementation Details}
\noindent\textbf{Datasets}. We train our model using the Owlii dataset\citep{2018_Owlii}, which comprises four sequences totaling 2400 frames, with each sequence lasting 20 seconds at 30 frames per second (FPS). To optimize training efficiency, we quantize the original 11-bit precision point cloud data to 10-bit. For model evaluation, we utilize the MPEG 8iVFB dataset \citep{2017_8iVFB}, consisting of four sequences with 1200 frames, where each sequence spans 10 seconds at 30 FPS. We adhere to the MPEG common test condition (CTC), setting the GoF size to 32. The first frame of each GoF is designated as an ``I" frame and encoded by PCGCv2\citep{2021_PCGCv2} with the same $\lambda$, while subsequent frames are labeled as ``P" frames and encoded by the proposed method.

\begin{table*}[t!]
  \centering
  \newcolumntype{C}{>{\centering\arraybackslash}X}
  \begin{tabularx}{\textwidth}{@{} >{\hsize=1.5\hsize}X *{9}{>{\hsize=.9444\hsize}C} @{}} 
    \toprule
    \multirow{2}{*}{Target bitrate}  & \multicolumn{9}{c}{Bitrate error $\Delta R\%$ } \\
    \cline{2-10}
    &  0.15 & 0.17& 0.19 & 0.21 & 0.23 & 0.25& 0.27 & 0.29 & \textbf{Average}\\
    \midrule
    Longdress & 0.07 & 0.25 & 0.23 & 0.28 & 0.33 & 0.42 & 0.53 & 0.32 & \textbf{0.30}\\
    Loot & 0.44 & 0.23 & 1.02 & 0.27 & 0.18 & 0.10 & 0.01 & 1.16 & \textbf{0.43}\\
    Redandblack & 1.12 & 0.67 & 1.43 & 0.21 & 0.69 & 0.16 & 1.00 & 0.22 & \textbf{0.69}\\
    Soldier & 0.37 & 0.42 & 0.38 & 0.04 & 0.06 & 0.10 & 0.12 & 0.03 & \textbf{0.19}\\
    \bottomrule
  \end{tabularx}
  \caption{Bitrate error comparison on 8iVFB dataset.}
  \label{tab:ratecontrol}
\end{table*}

\noindent\textbf{Network configuration}. Through numerous trials, we finalized our network configuration to include four coding routes ($K=4$), with a Lagrange multiplier list set to $[3,7,10,20]$. The training strategy follows the protocol established in Algorithm \ref{alg:joint_training}. The training is conducted on an NVIDIA A100 GPU, lasting for 100 epochs with the batch size of 2.

\noindent\textbf{Evaluation metrics}.
We measure bitrate in bits per point (bpp), calculated by dividing the total bitstream size by the number of input points. Distortion is assessed using point-to-point (D1) PSNR and point-to-plane (D2) PSNR, which quantify the reconstruction accuracy compared to the original point clouds. Rate control accuracy is evaluated by the bitrate error \(\Delta R=\frac{|R_{out}-R_{tar}|}{R_{tar}} \times 100\%\), and coding time complexity by the average coding time per frame \(T=\frac{T_{total}}{N_{frame}}\).

\noindent\textbf{Benchmark models}. 
To validate the effectiveness of our approach, we compare it against several PCC methods: rule-based V-PCC test model v18 \citep{2019_MPEG}, learning-based D-DPCC \citep{2022_DDPCC}, patchDPCC \citep{2024_patch_DPCC}, and PCGCv2 \citep{2021_PCGCv2}. Static PCC methods like PCGCv2 encode frames independently. The results for patchDPCC, whose source code is unavailable, are taken directly from the original paper. To ensure fairness, we maintain identical testing conditions to patchDPCC. Additionally, D-DPCC and PCGCv2 are retrained under the same conditions as the proposed method.

\subsection{Experimental Results}

\begin{figure*}[t!]
\centering
\includegraphics[width=1.0\textwidth]{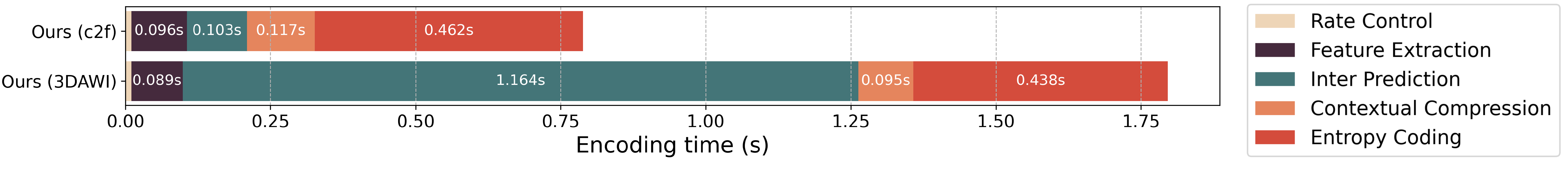}
\caption{Average encoding time comparison between the proposed coarse-to-fine (c2f) inter-frame prediction and the 3DAWI method on the \textit{Soldier} sequence. Both methods were controlled to achieve 0.20 bpp for fairness.}
\label{fig:ablation_coding_latency}
\end{figure*}

\begin{figure}[t!]
\centering
\includegraphics[width=0.45\textwidth]{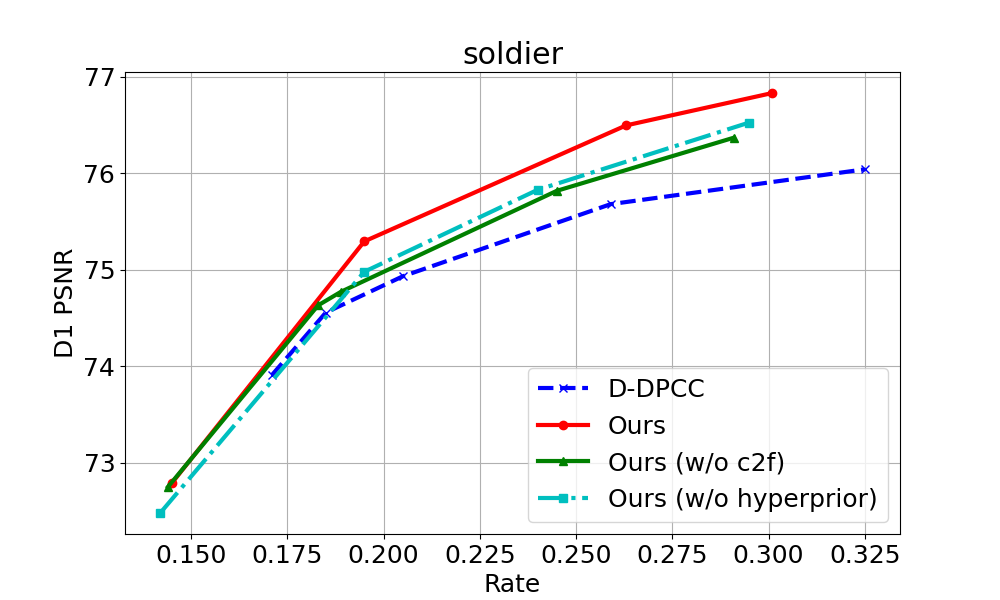}
\caption{Ablation study on RD performance: comparison of our method with and without the coarse-to-fine (c2f) module and hyperprior context.}
\label{fig:ablation_RD}
\end{figure}

\noindent\textbf{RD performance}. Figure \ref{fig:RD_Curve} shows the RD curves of our method on the 8iVFB dataset, illustrating superior RD performance in a single model compared to benchmark methods. Table \ref{tab:RDperformance} further details quantitative results. Our method outperforms rule-based V-PCC, achieving an average BD-Rate reduction of 62.439\% and a 4.096\,dB improvement in BD-PSNR. Against the advanced learning-based method patchDPCC, our method shows a BD-Rate reduction of 5.81\% and a 0.423\,dB increase in BD-PSNR.


\noindent\textbf{Coding time}. The coding times for various methods are compared in Table \ref{tab:complexity}, all evaluated under the same hardware and software environment using an Nvidia GeForce GTX 3090 GPU with CUDA 11.6. The coding time for V-PCC, which exceeds 50 seconds, is not listed as it operates on a CPU platform, contrasting with the GPU-based environment used for learning-based methods. Our method's maximum coding time per frame (via Route 3) is approximately 2.23 seconds, which is still 8.2\% faster than D-DPCC. As the computational load decreases, the coding time reduces significantly, with a 39.4\% reduction from Route 3 to Route 0, and it is 44.6\% faster than D-DPCC, demonstrating our method’s efficiency for real-time DPCC applications.

\noindent\textbf{Rate control accuracy}.
Our rate control module's accuracy is tested by setting target bitrates from 0.15 to 0.29 bpp in increments of 0.02 bpp. As shown in Table \ref{tab:ratecontrol}, our method consistently achieved close adherence to the set targets, with an average bitrate error of only 0.40\%, showcasing its precise rate control capabilities.

\subsection{Ablation Study}

\noindent\textbf{Coding latency}. To validate the low coding latency of our proposed coarse-to-fine inter-frame prediction, we conducted an ablation study by replacing it with an alternative inter-frame prediction method, KNN-based 3DAWI \citep{2022_DDPCC}, and measured the encoding time of each component. Experimental results on the \textit{Soldier} sequence are shown in Figure \ref{fig:ablation_coding_latency}. The decomposed encoding time reveals that the 3DAWI method occupies 64.8\% of the total encoding time. In contrast, our proposed coarse-to-fine prediction model requires only 8.8\% of the time needed by 3DAWI, accounting for just 13.1\% of the total encoding time. This substantial reduction is primarily due to the use of computationally efficient sparse convolution instead of the time-consuming KNN. Additionally, Figure \ref{fig:ablation_coding_latency} shows that the inference latency of the rate control module is negligible, which is approximately 0.01 seconds, making it an efficient solution for real-time rate control.

\noindent\textbf{RD performance}.
To demonstrate the RD improvements achieved by the coarse-to-fine inter-frame prediction and hyper-prior context, we conducted an ablation study substituting the former with 3DAWI and removing the latter. Ablation results, shown in Figure \ref{fig:ablation_RD}, indicate that our method's RD improvement is driven by both components: the coarse-to-fine inter-frame prediction provides a 3.4\% BD-Rate reduction, while the hyper-prior context contributes an additional 2.9\%. These findings highlight the significant contributions of both components in enhancing RD performance.

\section{Conclusion}

In this work, we introduce a novel DPCC framework that achieves variable bitrate and computational complexities with an efficient rate control mechanism. By coarse-to-fine inter-frame prediction, the receptive field is expanded for precise motion estimation and compensation. To meet target bitrates, the lightweight rate control module adaptively navigates point cloud frames through various coding routes, ensuring precise rate control. Experimental results further demonstrate the proposed method's RDCO effectiveness and rate control accuracy, offering a flexible solution for real-time, bitrate-constrained DPCC applications.

\section{Acknowledgments}


This work was supported by The Major Key Project of PCL (PCL\allowbreak2024A02), Natural Science Foundation of China (6227\allowbreak1013, 6203\allowbreak1013), Guangdong Provincial Key Laboratory of Ultra High Definition Immersive Media Technology (2024B1212010006), Guangdong Province Pearl River Talent Program (2021QN020708), Guangdong Basic and Applied Basic Research Foundation (2024A15150\allowbreak10155), Shenzhen Science and Technology Program (JCYJ202408131602\allowbreak02004, JCYJ20230807120808017).


\nocite{*}
\bibliography{aaai25}

\end{document}